\documentclass[aps,pra,twocolumn,showpacs]{revtex4}
\usepackage{color}
\usepackage{graphicx}
\usepackage{amsmath}
\usepackage{amsfonts}
\usepackage{amssymb}
\newcommand{\ket}[1]{|#1\rangle}
\newcommand{\bra}[1]{\langle #1|}
\newcommand{\Tr}{\text{Tr}}
\begin{document}
\today
\title{Non-Abelian geometric phases in a system of coupled quantum bits}
\author{Vahid Azimi Mousolou\footnote{Electronic address: vahid.mousolou@lnu.se}} 
\affiliation{Department of Physics and Electrical Engineering, Linnaeus University,
 391 82 Kalmar, Sweden}
\author{Erik Sj\"oqvist\footnote{Electronic address: erik.sjoqvist@kemi.uu.se}}
\affiliation{Department of Quantum Chemistry, Uppsala University, Box 518,
Se-751 20 Uppsala, Sweden}
\affiliation{Centre for Quantum Technologies, National University of Singapore,
3 Science Drive 2, 117543 Singapore, Singapore}
\begin{abstract}
A common strategy to measure the Abelian geometric phase for a qubit is to let it 
evolve along an `orange slice' shaped path connecting two antipodal points on the Bloch 
sphere by two different semi-great circles. Since the dynamical phases vanish for such paths, 
this allows for direct measurement of the geometric phase. Here, we generalize 
the orange slice setting to the non-Abelian case. The proposed method to measure the 
non-Abelian geometric phase can be implemented in a cyclic chain of four qubits with 
controllable interactions.  
\end{abstract}
\maketitle
\section{Introduction}
The geometric phase (GP), first discovered by Berry \cite{berry84} for adiabatic cyclic changes
of pure quantum states, has been generalized to a wide range of contexts, such as non-adiabatic \cite{aharonov87}, non-cyclic \cite{samuel88}, non-Abelian \cite{wilczek84,anandan88}, and 
mixed state \cite{sjoqvist00,tong04} evolution. This purely geometric object is manifested in various 
theoretical and experimental areas, such as in optics, condensed matter physics, and molecular 
physics, as well as in quantum field theory, and more recently in quantum computation \cite{bohm03,Chruscinski04}.

An essential ingredient when measuring GP is to find techniques to remove 
the effect of dynamical phases associated with the Hamiltonian of the system. One such method 
is based on that there are certain paths along which the dynamical phase vanishes. For two-level 
systems (qubits), the `orange slice' shaped path, formed by connecting two antipodal points on the 
Bloch sphere along two different semi-great circles, is associated with vanishing dynamical phase, 
which allows for direct measurement of GP. The orange slice shaped path has 
indeed been a common method to measuring the Abelian GP  
\cite{kwiat91,allman97,du03,rippe08,sponar10}. 

Here, we generalize the concept of orange slice shaped path to the non-Abelian GP 
in non-adiabatic evolution \cite{anandan88}. The generalization can be realized by combining 
a pair of pulsed interactions in a cyclic chain of four coupled qubits. The resulting non-Abelian 
GP can be used for universal non-adiabatic holonomic single-qubit gates. Such 
gates have recently been proposed in Ref. \cite{sjoqvist12} and experimentally realized in Refs. 
\cite{abdumalikov13,feng13}. The interactions is assumed to be controllable and of combined 
XY and Dzialoshinski-Moriya type. Typical physical systems to implementing the non-Abelian 
orange slice path and corresponding GP could be coupled quantum dots 
\cite{mousolou14}, atoms trapped in an optical lattice \cite{radic12}, and surface states of 
topological insulators \cite{zhu11}. 

The outline of the paper is as follows. In the next, section we review the basic idea of 
orange slice path for non-adiabatic evolution of a single qubit. We demonstrate how 
such a path can be implemented by applying an appropriate pair of pulses and how the 
resulting GPs can be used to realize a single-qubit phase shift gate. 
In Sec. \ref{sec:nonabelian}, we generalize the orange slice setting to the non-Abelian 
case. Here, the orange slice shaped path consists of pairs of geodesic segments in the 
Grassmann manifold describing the system. We further demonstrate how the resulting 
non-Abelian GP can be measured. The paper ends with 
the conclusions.

\section{Abelian setting}
\label{sec:abelian}
We first describe the realization of the orange slice shaped path in the Abelian case of a 
single qubit. Let $\sigma_x,\sigma_y$, and $\sigma_z$ be the standard Pauli operators 
and consider the Hamiltonian  
\begin{eqnarray}
H^{(\textrm{a})} (t) = 
\frac{1}{2} f(t) \left[ \cos (\phi) \sigma_x + \sin (\phi) \sigma_y \right] , 
\label{A-hamiltonian}
\end{eqnarray}
where $f(t)$ and $\phi$ are externally controllable parameters; $f(t)$ defines the 
`pulse area' $\alpha_t = \int_0^t f(t') dt'$ and the angle variable $\phi$ is assumed 
to be constant over each pulse. A spin-half particle interacting with an external magnetic 
field in the $xy$-plane or a polarized photon moving through a half wave plate, where 
the direction of the optical axis is given by $\phi$, are possible realizations of $H^{(\textrm{a})}$. 
Turning on $f(t)$ at $t=0$, the system is described by the time evolution operator ($\hbar = 1$ 
from now on)
\begin{eqnarray}
\mathcal{U}^{(\textrm{a})} (t,0) & = & e^{-i\int_0^t H^{(\textrm{a})} (t') dt'} 
\nonumber\\
 & = & \left( \begin{array}{cc}
\cos (\alpha_t /2) &-i e^{-i\phi} \sin (\alpha_t /2)  \\
-i e^{i\phi} \sin (\alpha_t /2) & \cos (\alpha_t /2) 
\end{array} \right) 
\label{a-u(t)}
\end{eqnarray}
expressed in the computational qubit basis $\{ \ket{0},\ket{1} \}$, where $\sigma_z \ket{n} = 
(1-2n)\ket{n}$, $n=0,1$. Here, $\ket{0}$ and $\ket{1}$ represent the north and south poles, 
respectively, of the Bloch sphere. 

\begin{figure}[h]
\centering
\includegraphics[scale=0.4]{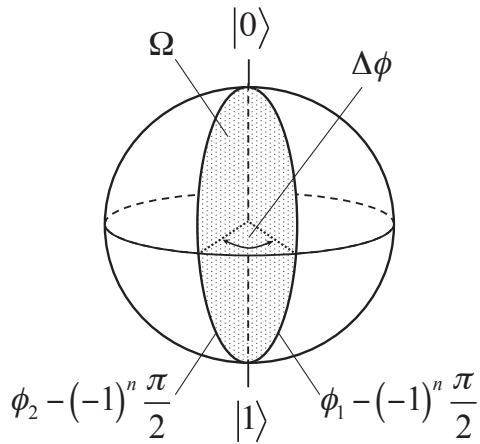}
\caption{Orange slice shaped path traced out by a qubit as it precesses around the Bloch sphere 
from one of the poles $\ket{n}$, $n=0,1$, on a semi-great circle to opposite pole, and then 
return on a different semi-great circle to the initial pole. The GP is proportional 
to the solid angle $\Omega = 2(\phi_2-\phi_1)$ enclosed by the orange slice shaped path.}
\label{fig:aosc}
\end{figure}

Suppose first the system is prepared at one of the poles $\ket{n}$, $n=0,1$, of the Bloch 
sphere. Turn on $f(t)$ and $\phi_1$ for the time interval $[0, \tau]$, then switch to $-f(t-\tau)$ 
and $\phi_2$ for the time interval $[\tau, 2\tau]$. We choose $\alpha_{\tau}=\int_0^{\tau} f(t') dt' 
= \pi$.  This evolves the system from the starting pole to the opposite pole along the semi-great 
circle on the Bloch sphere corresponding to the fixed angle $\phi_1-(-1)^{n} \frac{\pi}{2}$ and 
then back to the initial pole along different semi-great circle corresponding to the fixed angle 
$\phi_2-(-1)^{n} \frac{\pi}{2}$. In other words, this evolution takes the system around the orange 
slice shaped path on the Bloch sphere characterized by  pairs of azimuthal angles 
$\phi_1-(-1)^{n} \frac{\pi}{2}$ and $\phi_2-(-1)^{n}\frac{\pi}{2}$ depending on the initial 
state $\ket{n}$ of the qubit, as sketched in  Fig. \ref{fig:aosc}. The final time evolution operator 
is given by
\begin{eqnarray}
\mathcal{U}^{(\textrm{a})} (2\tau,0) & = & \mathcal{U}^{(\textrm{a})} (2\tau,\tau;\phi_{2})
\mathcal{U}^{(\textrm{a})} (\tau,0;\phi_{1}) 
\nonumber\\
 & = & \left( \begin{array}{cc}
 0 & ie^{-i\phi_{2}} \\
 ie^{i\phi_{2}} & 0 
\end{array} \right) 
\left( \begin{array}{cc}
 0 & -i e^{-i\phi_{1}} \\
 -ie^{i\phi_{1}} & 0 
\end{array} \right) 
\nonumber\\
&=& \left( \begin{array}{cc}
 e^{-i\Omega /2} & 0 \\
 0 & e^{i\Omega /2}  
\end{array} \right) ,
\label{eq:A-t-evolution}
\end{eqnarray}
where $\Omega = 2(\phi_2 - \phi_1) = 2\Delta \phi$ is the solid angle subtended by the 
two semi-great circles. The dynamical phases $-\int_0^t \bra{n} H^{(\textrm{a})} (t') 
\ket{n} dt'$, $n=0,1$, vanish along this evolution and hence the accumulated phases are 
purely geometric, defining the geometric phase shift gate
\begin{eqnarray}
\ket{n} \mapsto e^{-i(\frac{1}{2}-n)\Omega} \ket{n}, \ n=0,1. 
\end{eqnarray}
GP associated with this orange slice shaped path has been measured in several experiments   
\cite{kwiat91,allman97,du03,rippe08,sponar10}. 

\section{Non-Abelian generalization}
\label{sec:nonabelian} 
\subsection{Model system}
We now extend the above Abelian setting to the non-Abelian case. Note that while the Abelian 
GP appears in the evolution of a pure state, which constitutes a one-dimensional 
subspace of the full Hilbert space, the non-Abelian GP is a property of a 
higher-dimensional subspace \cite{wilczek84, anandan88}. Thus, we need more than a 
single qubit to achieve this. The general structure that we have in mind is described by a 
cyclic chain of four qubits with nearest-neighbor interaction described by the Hamiltonian 
\begin{eqnarray}
H =\frac{1}{2} F(t) \sum_{k=1}^4 \left( J_{k,k+1} R_{k,k+1}^{\textrm{XY}} + 
D_{k,k+1}^z R_{k,k+1}^{\textrm{DM}} \right) ,
\label{eq:4-qubit}
\end{eqnarray}
where $R_{k,k+1}^{\textrm{XY}} = \frac{1}{2} \left( \sigma_x^k \sigma_x^{k+1} + \sigma_y^k 
\sigma_y^{k+1} \right)$ and $R_{k,k+1}^{\textrm{DM}} = \frac{1}{2} \left( \sigma_x^k 
\sigma_y^{k+1} - \sigma_y^k \sigma_x^{k+1} \right)$ are XY and Dzialochinski-Moriya (DM)
terms with coupling strengths $J_{k,k+1}$ and $D_{k,k+1}^z$, respectively. $F(t)$ 
turns on and off all qubit interactions simultaneously. The cyclic nature of the qubit chain 
is reflected in the boundary conditions $J_{4,5} R_{4,5}^{\textrm{XY}} = J_{4,1} R_{4,1}^{\textrm{XY}}$ 
and $D_{4,5}^z R_{4,5}^{\textrm{DM}} = D_{4,1}^z R_{4,1}^{\textrm{DM}}$. 

The Hamiltonian in Eq. (\ref{eq:4-qubit}) preserves the single-excitation subspace 
$\mathcal{H}_{\text{eff}}$ of the four qubits spanned by the following ordered basis
\begin{eqnarray}
\mathcal{B}=\{ \ket{1000},\ket{0010},\ket{0100},\ket{0001} \},
\end{eqnarray}
where $\ket{1000}$, say, stands for $\ket{1}_1\ket{0}_2\ket{0}_3\ket{0}_4$, 
$\ket{0}_{k}\ (\ket{1}_{k})$ being the eigenvector of the $z$ component of the Pauli operator 
$\sigma_z^k$ at site $k$ corresponding to the eigenvalue $1\ (-1)$. In the basis $\mathcal{B}$, 
the Hamiltonian takes the form 
\begin{eqnarray}
H^{(\textrm{na})} = \frac{1}{2}F(t) \left( \begin{array}{rr}
0 & T \\
T^{\dagger} & 0 
\end{array} \right) ,
\label{eq:hamiltonian}
\end{eqnarray}
where  
\begin{eqnarray}
T = \left( \begin{array}{rr}
J_{12} - iD_{12}^z & J_{41} + iD_{41}^z  \\
J_{23} + iD_{23}^z  & J_{34} - iD_{34}^z  
\end{array} \right) = U S V^{\dagger} . 
\end{eqnarray}
Here, $U,V$, and $S$ are the unitary and diagonal positive parts in the singular-value 
decomposition of $T$.  For simplicity, we assume $S>0$.  

There are different physical realizations of the Hamiltonian in Eq. (\ref{eq:hamiltonian}). First, 
it describes the single excitation subspace of a cyclic chain of four coupled quantum dots with 
double occupancy of each dot being prevented by strong Hubbard-repulsion terms \cite{mousolou14}. 
Secondly, $H^{(\textrm{na})}$ is relevant to a square optical lattice of two-level atoms with 
synthetic spin orbit coupling localized at each lattice site allowing for the desired combination of 
XY and DM interactions, by suitable parameter choices \cite{radic12}. Finally, the 
Ruderman-Kittel-Kasuya-Yosida interaction in three-dimensional topological insulators  
may be used to obtain the XY and DM interaction terms in $H^{(\textrm{na})}$ \cite{zhu11}. 

The Hamiltonian in Eq. (\ref{eq:hamiltonian}) splits the effective state space $\mathcal{H}_{\text{eff}}$ 
into two orthogonal subspaces, i.e.,
\begin{eqnarray}
\mathcal{H}_{\text{eff}}=M_{0}\oplus M_{1},
\end{eqnarray}
where $M_{0} = \textrm{Span} \{\ket{1000} , \ket{0010} \}$ and $M_{1} = \textrm{Span} \{ \ket{0100}, \ket{0001} \}$.
This implies that in the basis $\mathcal{B}$ the time evolution operator splits into $2 \times 2$ blocks 
according to \cite{mousolou14}
\begin{widetext}
\begin{eqnarray}
\mathcal{U}^{(\textrm{na})} (t,0) = \left( \begin{array}{cc} 
U \cos \left( \alpha_t S/2 \right) U^{\dagger}  & -i U \sin \left( \alpha_t S/2 \right) V^{\dagger} \\ 
-i V \sin \left( \alpha_t S/2 \right) U^{\dagger} &  V \cos \left( \alpha_t S/2 \right) V^{\dagger}  
\end{array} \right),
\label{eq:na-u(t)}
\end{eqnarray}
\end{widetext}
where $\alpha_t = \int_0^t F(t') dt'$. 

The non-Abelian orange slice path is realized by first applying a pulse over 
$[0,\tau]$ with $F(t)$ and $T_1=U_1 S_1 V_1^{\dagger}$, followed by a pulse 
over $[\tau',\tau' + \tau'']$ with $-\tilde{F}(t-\tau')$ and $T_2=U_2 S_2 V_2^{\dagger}$. We 
assume that $\tau' + \tau'' > \tau' > \tau$ and that the Hamiltonian is completely turned off 
on $[\tau,\tau']$. Note that the size of the time gap $\tau' - \tau$ is only restricted by 
errors, such as parameter noise and decoherence, and can therefore be arbitrarily long 
in the ideal error-free case. By choosing parameters such that 
$\cos \left( \alpha_{\tau} S_1/2 \right) = \cos \left( \tilde{\alpha}_{\tau''} S_2/2 \right) = 0$, 
$\sin \left( \alpha_{\tau} S_1/2 \right) = Z^{p_1}$, and $\sin \left( \tilde{\alpha}_{\tau''} S_2/2 
\right) = Z^{p_2}$, where $\tilde{\alpha}_t = \int_0^t \tilde{F}(t') dt'$, $p_1,p_2=0,1$, and 
$Z = \textrm{diag} \{ 1,-1 \}$, we obtain the final time evolution operator 
\begin{widetext}
\begin{eqnarray} 
\mathcal{U}^{(\textrm{na})} (\tau+\tau'',0) & = & \mathcal{U}^{(\textrm{na})} (\tau'+\tau'',\tau';T_2) 
\mathcal{U}^{(\textrm{na})} (\tau,0;T_1) = 
\left( \begin{array}{cc} 
0  & i U_2 Z^{p_2} V_2^{\dagger} \\ 
i V_2 Z^{p_2} U_2^{\dagger} &  0  
\end{array} \right)
\left( \begin{array}{cc} 
0  & -i U_1 Z^{p_1} V_1^{\dagger} \\ 
-i V_1 Z^{p_1} U_1^{\dagger} &  0  
\end{array} \right) 
\nonumber \\ 
 & = & \left( \begin{array}{cc} 
U_2 Z^{p_2} V_2^{\dagger} V_1 Z^{p_1} U_1^{\dagger} & 0\\ 
0 & V_2 Z^{p_2} U_2^{\dagger} U_1 Z^{p_1} V_1^{\dagger}
\end{array} \right) 
\end{eqnarray}
\end{widetext}
expressed with respect to the basis $\mathcal{B}$.

\begin{figure}[h]
\centering
\includegraphics[scale=0.28]{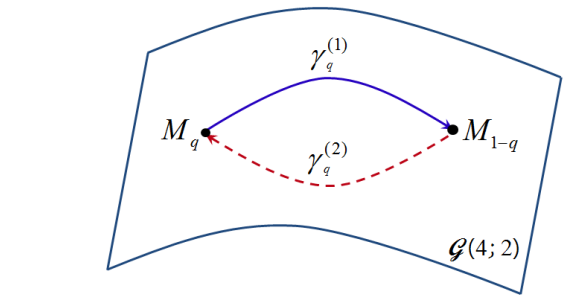}
\caption{Schematic picture of orange slice shaped paths $C_{q}=\gamma_{q}^{(1)}*\gamma_{q}^{(2)}, 
\ q=0,1,$ in the Grassmannian $\mathcal{G}(4;2)$. $\gamma_{q}^{(1)}$ and $\gamma_{q}^{(2)}$ 
are the two geodesic edges of the orange slice shaped path $C_{q}$, which connect the two 
orthogonal poles $M_0$ and $M_1$ in $\mathcal{G}(4;2)$.}
\label{fig:naosp}
\end{figure}

By considering the evolution of the orthogonal subspaces $M_{0}$ and $M_{1}$ in each pulse, 
one may notice that the two unitaries $U(C_{0}) = U_2 Z^{p_2} V_2^{\dagger} V_1 Z^{p_1} 
U_1^{\dagger}$ and $U(C_{1}) = V_2 Z^{p_2} U_2^{\dagger} U_1 Z^{p_1} V_1^{\dagger}$ are 
purely geometric since the Hamiltonian $H^{(\textrm{na})}$ vanishes on $M_{0}$ and $M_{1}$ 
separately \cite{anandan88}. In fact, for each $q=0,1$, $U(C_{q})$ is the non-adiabatic non-Abelian 
GP associated with the evolution of the subspace $M_{q}$ along the orange slice 
shaped path $C_{q}$ via the midpoint $M_{1-q}$ back to itself, as depicted in Fig. \ref{fig:naosp}. 
Thus, the time evolution operator $\mathcal{U}^{(\textrm{na})} (\tau'+\tau'',0)$ is fully determined 
by the pair of closed paths $C_{0}$ and $C_{1}$ in the Grassmannian $\mathcal{G}(4;2)$, i.e., the 
space of two-dimensional subspaces of a four-dimensional Hilbert space \cite{bengtsson06}.

In the following we will justify the orange slice nature of the path $C_{q}$ and propose a 
measurement mechanism of the associated non-Abelian GP $U(C_{q})$.

\subsection{Geometric interpretations}
The path $C_q$ is formed of two geodesic paths $\gamma_q^{(1)}$ and $\gamma_q^{(2)}$ in 
$\mathcal{G} (4;2)$. To see this, consider the Stiefel manifold $\mathcal{S}(4;2)$, which is 
the space of $2$-frames in the effective four-dimensional Hilbert space $\mathcal{H}_{\text{eff}}$. 
There is a natural projection
\begin{eqnarray}
\Pi\ :\ \mathcal{S}(4;2)\longrightarrow \mathcal{G}(4;2), 
\end{eqnarray} 
which takes each frame to the corresponding subspace spanned by that frame. 
For each $q=0, 1$ and $l=1,2$ we may introduce the new orthonormal basis vectors of $\mathcal{H}_{\text{eff}}$ as 
\begin{eqnarray}
\ket{e_{1}^{(l, q)}}&=&\Lambda(q)\ket{(1-q)q00},
\nonumber \\
\ket{e_{2}^{(l, q)}}&=&\Lambda(q)\ket{00(1-q)q}, 
\nonumber \\
\ket{e_{3}^{(l, q)}}&=&(-1)^{l}\Lambda(q)\ket{q(1-q)00}, 
\nonumber \\
\ket{e_{4}^{(l, q)}}&=&(-1)^{l}\Lambda(q)\ket{00q(1-q)},
\end{eqnarray}
where the unitary operator $\Lambda(q)$ has the form
\begin{eqnarray}
\Lambda(q)= \left( \begin{array}{rr}
i^{q}U&0\ \ \ \ \\
0\ \ &\ \  i^{1-q}V
\end{array} \right)  
\end{eqnarray}
in the basis $\mathcal{B}$.  With the above notations, one may notice that  $\Pi \left[ 
\ket{e_{1}^{(l, q)}}, \ket{e_{2}^{(l, q)}} \right] = M_{q}$ and $\Pi \left[ \ket{e_{3}^{(l, q)}}, 
\ket{e_{4}^{(l, q)}} \right] = M_{1-q}$.

In the non-Abelian case, subspaces $M_{0}$ and $M_{1}$ play the same role as the two 
opposite poles in the Abelian case. The non-Abelian evolution takes the system initially 
prepared in the subspace $M_{q}$ to its orthogonal complement $M_{1-q}$ along the 
geodesic
\begin{widetext}
\begin{eqnarray}
\gamma_{q}^{(1)}\ :\ [0,\tau]\ni t\longrightarrow & & \Pi \left[ \cos \left( \frac{\alpha_t}{2} s_{1;1} \right)  
\ket{e_{1}^{(1, q)}} + \sin \left( \frac{\alpha_t}{2} s_{1;1} \right)  \ket{e_{3}^{(1, q)}}, \right. 
\nonumber \\ 
 & & \left. 
\cos \left( \frac{\alpha_t}{2} s_{1;2} \right)  \ket{e_{2}^{(1, q)}} + \sin \left( \frac{\alpha_t}{2} s_{1;2} 
\right)  \ket{e_{4}^{(1, q)}} \right],
\label{eq: geodesic1}
\end{eqnarray} 
\end{widetext}
in $\mathcal{G}(4;2)$, where $S_1=\text{diag}\{s_{1;1}, s_{1;2}\}>0$ \cite{zhou98}. 
This is followed by evolving the subspace $M_{1-q}$ back to the initial subspace 
$M_{q}$ along a different geodesic in $\mathcal{G}(4;2)$ given by
\begin{widetext}
\begin{eqnarray}
\gamma_{q}^{(2)}\ :\ [\tau',\tau'+\tau'']\ni t\longrightarrow & & 
\Pi \left[ \cos \left( \frac{\tilde{\alpha}_t}{2} s_{2;1} \right) \ket{e_{1}^{(2, 1-q)}} + 
\sin \left( \frac{\tilde{\alpha}_t}{2} s_{2;1} \right)  \ket{e_{3}^{(2, 1-q)}}, \right. 
\nonumber \\ 
 & & \left. \cos \left( \frac{\tilde{\alpha}_t}{2} s_{2;2} \right)  \ket{e_{2}^{(2, 1-q)}}  + 
\sin \left( \frac{\tilde{\alpha}_t}{2} s_{2;2} \right)  \ket{e_{4}^{(2, 1-q)}} \right], 
\nonumber\\
\label{eq: geodesic2}
\end{eqnarray} 
\end{widetext}
where $S_2=\text{diag}\{s_{2;1}, s_{2;2}\}>0$. In fact for each $q=0,1$, the geodesics 
$\gamma_{q}^{l}, \ l=1,2,$ form the two edges of the orange slice shaped path $C_q$, 
which connect the two orthogonal subspaces $M_{0}$ and $M_{1}$ in $\mathcal{G}(4;2)$.

The orange slice nature of the path $C_{q}$ in $\mathcal{G}(4;2)$ and the resulting non-Abelian 
GP may be more transparent in the following intuitive picture. We first examine 
the corresponding paths leading to the Abelian GPs in Eq. (\ref{eq:A-t-evolution}). 
Let us consider the initial state $\ket{0}$, which evolves as  
\begin{eqnarray}
\ket{0} & \mapsto & \mathcal{U}^{(\textrm{a})} (t,0) \ket{0} = \ket{\psi_0 (t)}
\nonumber \\ 
 & = &  \cos (\alpha_t /2) \ket{0} -i e^{i\phi} \sin (\alpha_t /2) \ket{1} . 
\end{eqnarray}
This vector rotates at the same constant angle around the fixed vectors 
\begin{eqnarray}
\ket{\pm (\phi)} = \frac{1}{\sqrt{2}} \left( \ket{0} \pm e^{i\phi} \ket{1} \right)
\end{eqnarray}
in the sense that the fidelity is time independent, viz., $\left| \langle \pm (\phi) 
\ket{\psi_0 (t)} \right| = \frac{1}{\sqrt{2}}$. Thus, $\ket{\psi_0 (t)}$ is mutually 
unbiased with respect to each of $\ket{\pm (\phi)}$. The vectors $\ket{\pm (\phi)}$ 
are eigenvectors of the fixed Pauli operator ${\bf n} \cdot \boldsymbol{\sigma} = 
\cos (\phi) \sigma_x + \sin (\phi) \sigma_y$, ${\bf n}$ being the rotation axis of 
the Bloch vector for each semi-great circle and $\boldsymbol{\sigma} = (\sigma_x,
\sigma_y,\sigma_z)$. 

The above picture translates to the non-Abelian setting as follows. We consider the evolution 
of $M_{0}$  in $\mathcal{G}(4;2)$. This evolution can be 
viewed as a rotating complex $2$-plane in the four-dimensional complex vector space 
$\mathcal{H}_{\text{eff}}$. This plane is spanned  by the frame $\{ \mathcal{U}^{(\textrm{na})} 
(t,0) \ket{\mu_1}, \mathcal{U}^{(\textrm{na})} (t,0) \ket{\mu_2} \}$, where $\ket{\mu_1}$ and 
$\ket{\mu_2}$ can be any pair of orthonormal vectors spanning $M_{0}$. In particular, the 
vectors $\ket{\mu_1} = \Lambda(0)\ket{1000}$ and $\ket{\mu_2} = \Lambda(0)\ket{0010}$ 
defines the $2$-frame  
\begin{eqnarray}
\ket{\mu_1 (t)} = \Lambda(0)\left[\cos \left( \frac{\alpha_t}{2} s_1 \right)\ket{1000}
-\sin \left( \frac{\alpha_t}{2} s_1 \right)\ket{0100}\right] , 
\nonumber\\
\ket{\mu_2 (t)}  = \Lambda(0)\left[\cos \left( \frac{\alpha_t}{2} s_2 \right)\ket{0010}
- \sin \left( \frac{\alpha_t}{2} s_2 \right)\ket{0001}\right], 
\nonumber\\
\end{eqnarray}
where $S = \textrm{diag} \{ s_1,s_2 \} > 0$. Thus, $P(t) = \ket{\mu_1 (t)} 
\bra{\mu_1 (t)} + \ket{\mu_2 (t)} \bra{\mu_2 (t)}$ projects onto the rotating complex 
$2$-plane $M(t)$. 

Now, there are fixed two-dimensional subspaces $M_{\pm}$ with corresponding projection 
operators $P_{\pm}$ that satisfy the fidelity relations 
\begin{eqnarray}
\Tr \left[ P_{\pm}P (t) \right] = 1,
\end{eqnarray}
being again the same for $\pm$ and independent of time. Explictly, a $2$-frame spanning 
$M_{\pm}$ is 
\begin{eqnarray}
\ket{\psi_{\pm}} & = & \frac{1}{\sqrt{2}} \Lambda(0)\left( \ket{1000}  \pm i\ket{0100}  \right) , 
\nonumber \\ 
\ket{\psi_{\pm}^{\perp}} & = & \frac{1}{\sqrt{2}}\Lambda(0) \left( \ket{0010}  \pm i\ket{0001}  \right) ,  
\end{eqnarray}
which can be used to prove that the two subspaces $M (t)$ and $M_{\pm}$ are mutually unbiased. 
Similar scenarios hold for evolutions initiated at $\ket{1}$ and $M_{1}$. Thus, in analogy with the 
above Abelian case, we may understand the orange slice nature of $C_q$ as a rotation of the 
complex $2$-plane around some fixed subspaces.

Notice that $U_l Z^{p_l} V_l^{\dagger}$ is a U(2) transformation and therefore can be put 
on the form $e^{-i\chi_l} e^{-i\varphi_l {\bf n}_l \cdot {\bf X}/2}$, where $\chi_l$ and $\varphi_l$ 
are real-valued, ${\bf n}_{l}$ is a real unit vector, and ${\bf X}$ is the vector of standard Pauli 
matrices acting on $M_0$ or $M_1$. Thus, the GPs may be written 
\begin{eqnarray}
U(C_{0}) & = & e^{-i{\bf \Delta\chi}} e^{-i\varphi_2 {\bf n}_2 \cdot {\bf X}/2} 
e^{i\varphi_1 {\bf n}_1 \cdot {\bf X}/2} ,
\nonumber \\
U(C_{1}) & = & e^{i{\bf \Delta\chi}} e^{i\varphi_2 {\bf n}_2 \cdot 
{\bf X}/2} e^{-i\varphi_1 {\bf n}_1 \cdot {\bf X}/2} ,
\label{eq:nongp}
\end{eqnarray} 
where $\Delta\chi=\chi_2 - \chi_1$. 

The non-Abelian contributions $e^{-i\varphi_2 {\bf n}_2 \cdot {\bf X}/2} 
e^{i\varphi_1 {\bf n}_1 \cdot {\bf X}/2}$ and $e^{i\varphi_2 {\bf n}_2 \cdot {\bf X}/2} 
e^{-i\varphi_1 {\bf n}_1 \cdot {\bf X}/2}$ represent twists caused by the non-trivial 
geometry of $\mathcal{G}(4;2)$. The non-Abelian nature is apparent from the fact that the 
two factors in each GP do not commute when ${\bf n}_2 \neq {\bf n}_1$. The 
unitary action on $M_{q}$, that is induced by $C_{q}$, is universal since the SU(2) 
parameters $\varphi_1,\varphi_2,{\bf n}_1$, and ${\bf n}_2$ can in principle be chosen
independently. Thus, this unitary action serves as a universal gate on a single qubit 
encoded in $M_{q}$. On the other hand, the Abelian parts $e^{\pm i {\bf \Delta\chi}}$ 
are global phases and therefore unimportant for such single qubit  gate operations.
 
\subsection{Meaurement scheme}
The non-Abelian GPs $U(C_q)$ can be measured by applying unitary operators 
$W_q$ acting on the subspaces $M_q$ immediately after the realization of the orange slice path. 
This results in the unitary transformation $W_0 U(C_0) \oplus W_1 U(C_1)$. For an input 
state $\ket{\psi} \in \mathcal{H}_{\text{eff}}$, the survival probability reads 
\begin{eqnarray}
p  =  \left| \bra{\psi}W_0 U(C_{0})\oplus W_1 U(C_{1}) \ket{\psi} \right|^2 
\leq  1
\end{eqnarray}
with equality for all $\ket{\psi}$ when $W_{0}^{\dagger} \oplus W_{1}^{\dagger} = 
U(C_{0})\oplus U(C_{1})$ up to an overall U(1) phase factor. In this way, the non-Abelian 
SU(2) part of $U(C_{q})$ can be measured by varying $W_{q}$ until maximum is reached. 

We now describe how this scheme can be implemented in a cyclic chain of four coupled 
quantum dots at half-filling \cite{mousolou14, mousolou13}. Appropriate XY and DM terms can 
be designed in this system by utilizing the interplay between electron-electron repulsion 
and spin-orbit interaction. In this way, $U(C_q)$ and $W_q$ can be realized in the invariant 
four dimensional subspace, our $\mathcal{H}_{\text{eff}}$, spanned by the local single spin 
flip states $\ket{\!\!  \downarrow \uparrow \uparrow \uparrow}, \ldots, \ket{\! \! \uparrow 
\uparrow \uparrow \downarrow}$ of the electrons. 

The non-Abelian GPs $U(C_q)$ are realized in $\mathcal{H}_{\text{eff}}$ by 
turning on and off appropriate nearest-neighbor interactions, which results in an effective 
Hamiltonian of the form given by Eq. (\ref{eq:hamiltonian}). The variabe unitary operators 
$W_q$ should be $2 \times 2$ diagonal blocks in $\mathcal{H}_{\text{eff}}$, which is 
achieved by turning on and off the next-nearest-neighbor interactions as described 
by the Hamiltonan
\begin{eqnarray}
h & = & f(s)[ \sum_{k=1}^2 \left( J_{k,k+2} R_{k,k+2}^{\textrm{XY}} + 
D_{k,k+2}^z R_{k,k+2}^{\textrm{DM}} \right) 
\nonumber \\ 
 & & + E (Z_1 + Z_2)] . 
\end{eqnarray} 
To realize full variablity in $W_q$, we have added the term $E (Z_1 + Z_2)$ with $Z_1 = 
\ket{\!\!  \downarrow \uparrow \uparrow \uparrow} \bra{\downarrow \uparrow \uparrow 
\uparrow \!\!} - \ket{\!\!  \uparrow \uparrow \downarrow \uparrow} \bra{\uparrow \uparrow 
\downarrow \uparrow \!\!}$ and $Z_2 = \ket{\!\!  \uparrow \downarrow 
\uparrow \uparrow} \bra{\uparrow \downarrow \uparrow \uparrow \!\! } - 
\ket{\!\!  \uparrow \uparrow \uparrow \downarrow} \bra{\uparrow \uparrow 
\uparrow \downarrow \!\! }$ corresponding to a local energy shift of the first and second 
sites relative the third and fourth site (for instance by applying an inhomogeneous 
magnetic field over the four-dot system). In the ordered basis $\mathcal{B}=\{\ket{\!\!  \downarrow \uparrow \uparrow \uparrow}, \ket{\!\!  \uparrow \uparrow \downarrow \uparrow}, \ket{\!\!  \uparrow \downarrow \uparrow \uparrow}, \ket{\! \! \uparrow 
\uparrow \uparrow \downarrow}\}$ of the single spin flip subspace, $h =  f(s)
T_0 \oplus T_1$ with the $2\times 2$ blocks
\begin{eqnarray}
T_0 & = & \left( \begin{array}{cc}
E & J_{13} + iD_{13}^z  \\
J_{13} - iD_{13}^z  & -E 
\end{array} \right) , 
\nonumber \\ 
T_1 & = & \left( \begin{array}{cc}
E & J_{24} + iD_{24}^z  \\
J_{24} - iD_{24}^z  & -E 
\end{array} \right) . 
\end{eqnarray}
The variable unitary $W_0 \oplus W_1$ is generated by $h$ and takes the desired block-diagonal 
form $W_0 \oplus W_1 = e^{-ib_{t} T_0}\oplus e^{-ib_{t} T_1}$ with $b_t = \int_0^t f(s) ds$ the 
`pulse area'. Thus, $e^{-ib_t T_0}$ and $e^{-ib_t T_1}$ are SU(2) operators that can be fully 
varied by changing the parameters $J_{k,k+1},D_{k,k+1}$, and $E$.

To measure the non-Abelian GPs, prepare first an appropriate initial state in the 
four-dot system by polarizing the spins along the $z$ direction by an external magnetic field 
followed by a single spin flip induced by a local magnetic field in the $x$ direction at one of 
the sites \cite{grinolds11}. Suppose we apply the spin flip to the first site leading to the initial 
spin state $\ket{\psi} = \ket{\!\!  \downarrow \uparrow \uparrow \uparrow} \in 
\mathcal{H}_{\text{eff}}$ of the four electrons. In this way, a measurement of $U(C_{0})$ can 
be performed by implementing sequentially $\mathcal{U}^{(\textrm{na})}(\tau,0; T_{1})$, 
$\mathcal{U}^{(\textrm{na})} (\tau' + \tau'', \tau'; T_{2})$, and $e^{-ib_{t} h}$. $U(C_{0})$ can 
be measured by varying the parameters $J_{13}, D_{13}, E$ and $b_{t}$ until the probability 
$p = \left| \bra{\downarrow \uparrow \uparrow \uparrow\!\!  }e^{-ib_{t} T_{0}} U(C_{0}) 
\ket{\!\!  \downarrow \uparrow \uparrow \uparrow} \right|^2$ reaches its maximum at 
$e^{ib_{t} T_{0}} = U(C_{0})$.

\section{Conclusions}
We have developed a non-Abelian generalization of the concept of orange slice shaped paths 
to allow for direct measurement of the non-Abelian GP in non-adiabatic 
evolution. The geometric interpretation of this non-Abelian GP is quite 
different from that in terms of the solid angle enclosed on the Bloch sphere of its Abelian 
counterpart. Instead, the orange slice nature of the path underlying the non-Abelian GP 
is associated with pairs of geodesics in the Grassmannian manifold $\mathcal{G}(4;2)$, i.e., the 
space of two-dimensional subspaces of the system's effective four-dimensional Hilbert 
space. 

The proposed method to measuring the non-Abelian GP can be 
implemented in a cyclic chain of four qubits that can be realized in different physical settings, 
such as in quantum dot, optical lattice, and topological insulator architectures. The realizations 
and their read-out require interactions that can simultaneously be turned on and off in a 
controlled way. 

The present work suggests an alternative way to reach closed paths in the Grassmannian 
$\mathcal{G}(4;2)$, which is the main ingredient in achieving universal holonomic quantum 
gates proposed in \cite{mousolou14} to perform fault tolerant quantum information processing. 
The orange slice shaped loops considered here are encouraged by a common experimental 
method for measuring the Abelian GP  \cite{kwiat91,allman97,du03,rippe08,sponar10}. 
This certain type of closed paths is substantially different from those considered in Ref. 
\cite{mousolou14} in that the orange slice shaped loops are accomplished by applying 
sequentially two different pulses while the loops in Ref. \cite{mousolou14} are results of 
single pulses. The orange slice technique has the advantage that it allows for 
arbitrary SU(2) holonomic transformations in a single-loop scenario, while the scheme 
in Ref. \cite{mousolou14} can achieve this only by combining at least two loops. 

It is important that different approaches and schemes for holonomic quantum computation 
be thoroughly explored and compared with one another. This would help to optimize the setup 
with respect to robustness and, at the same time, to make it accessible experimentally and 
amenable to external manipulation. Such a setup would be essential to construct scalable,
compact, and reproducible building blocks of quantum computers.  This would also further 
improve our understanding of the relation between the abstract theoretical objects, such as 
the geometry of the Grassmannian manifolds, and practically observed quantum phenomena.

\section*{ACKNOWLEDGMENTS}
V.A.M. is supported by Department of Physics and Electrical Engineering at Linnaeus 
University (Sweden) and by the National Research Foundation (VR). E.S. acknowledges 
support from the National Research Foundation and the Ministry of Education (Singapore).

\end{document}